\documentclass{ifacconf}

\usepackage{graphicx}      
\usepackage{natbib}        
\usepackage{color}

\usepackage{amsfonts}
\usepackage{amsmath}

\newcommand{\randvar}[1]{\uppercase{#1}} 
\newcommand{\pce}[1]{\textsf{#1}} 
\newcommand{\upbd}[2][]{{#2}_{#1max}} 
\newcommand{\lowbd}[2][]{{#2}_{#1min}} 

\newcommand{\mbb}[1]{\mathbb #1}

\newcommand{\mcl}[1]{\mathcal #1}

\newtheorem{definition}{Definition}

\newtheorem{example}{Example}

\begin{document}
\begin{frontmatter}

\title{A Simulation Study on Turnpikes in Stochastic LQ Optimal Control
} 


\author[First]{Ruchuan Ou} 
\author[Second]{Michael Heinrich Baumann} 
\author[Second]{Lars Gr\"une}
\author[First]{Timm Faulwasser}

\address[First]{Institute for Energy Systems, Energy Efficiency and Energy Economics, TU Dortmund University, Germany (e-mail: timm.faulwasser@ieee.org).}
\address[Second]{Mathematical Institute, Universität Bayreuth, 
   Bayreuth, Germany (e-mail: $\{$michael.baumann,lars.gruene$\}$@uni-bayreuth.de)}

\begin{abstract}                
This paper presents a simulation study on turnpike phenomena in stochastic optimal control problems. We employ the framework of Polynomial Chaos Expansions (PCE) to investigate the presence of turnpikes in stochastic LQ problems. Our findings indicate that turnpikes can be observed in the evolution of PCE coefficients as well as in the evolution of statistical moments. Moreover, the turnpike phenomenon can be observed for optimal realization trajectories and with respect to the optimal stationary distribution. 
Finally, while adding variance penalization to the objective alters the turnpike, it does not destroy the phenomenon. 
\end{abstract}


\end{frontmatter}

\section{Introduction}
The last decade has seen substantial progress in terms of optimal and predictive control. This includes the analysis of stochastic MPC for set-point stabilization and the understanding of deterministic economic MPC schemes, wherein the objective is more general than a penalization of the distance to a given set-point. 

A crucial point in the analysis of economic MPC is the interplay of turnpike and dissipativity properties. The former refers to the phenomenon that for many Optimal Control Problems (OCPs), the optimal solutions for varying horizon and varying initial conditions are structurally similar. More precisely, the turnpike phenomenon refers to the fact that in the middle part of the horizons the optimal solutions stay within a neighborhood of the optimal steady state, see~\citep{Dorfman58,Mckenzie76,Carlson91} for classical references and \citep{Trelat15a,tudo:faulwasser20g,Stieler14a} for more recent results. It is worth to be noted that the research on turnpike properties of OCPs originated in economics. 
The analysis of the interplay between turnpike and dissipativity notions of OCPs has been investigated in a number of papers; indeed it can be shown that under mild assumptions the turnpike property is equivalent to a certain strict dissipativity notion~\citep{Gruene16a,epfl:faulwasser15h}. Moreover, this close relation can be exploited in the analysis of economic MPC schemes, see~\citep{kit:faulwasser18c} for a recent overview. 
However, when it comes to economic MPC under uncertainties much less has been done in terms of analysis---see \citep{Bayer16a}---despite the fact that in the economics literature a number of investigations on turnpike properties in stochastic problems have been conducted~see \citep{Mari89,KolY12}. 

Regarding numerical computations with stochastic uncertainties, there has been recent interest into Polynomial Chaos Expansions (PCEs). The core idea of PCE is that a random variable can be modeled as an $L_2$ function in an appropriate Hilbert space and that in this space a polynomial basis can be used to parametrize the random variable by deterministic coefficients. The idea dates back to \cite{Wiener38}. In recent years, PCE methods have been subject to renewed interest and have been widely investigated for uncertainty quantification~\citep{Sullivan2015}. While in principle the number of terms in the series expansion is infinite, numerical implementation requires truncation. Recently, it was shown that for polynomial explicit mappings the truncation from applying Galerkin projection to the first $L+1$ basis functions can be characterized in closed form, which enables to quantify the error and to choose sufficiently many basis functions such that the error vanishes~\citep{kit:muehlpfordt18a}. In systems and control, PCE has been used in a number of papers, e.g., in~\citep{Paulson14a,Mesbah15a,Kim2013}.  The PCE approach is also considered for uncertainty quantification in electrical power systems~\citep{kit:muehlpfordt17b} and gas networks~\citep{Gerster19a}. A major advantage of the PCE framework is that it allows the consideration of a large class of non-Gaussian random variables with finite variance. 

The present note conducts a simulation study on turnpike properties in stochastic Linear-Quadratic (LQ) OCPs. Specifically, we employ the \texttt{PolyChaos.jl} package by \cite{tudo:muehlpfordt19c} to solve example problems. The contribution is to demonstrate that in understanding turnpike properties in stochastic LQ OCPs, one needs to compare the stochastically optimal steady state with the optimal distribution in the middle of the optimization horizon. Moreover, our numerical experiments show that the deterministic PCE coefficients of the state and input variables also exhibit a turnpike phenomenon. Indeed one can also observe the turnpike phenomenon if the disturbance realization sequence is identical for different initial conditions. Finally, we also show that the phenomenon is robust under consideration of variance penalization in the objective. Our results illustrate the prospect of  systematically using stochastic turnpikes in the analysis of stochastic MPC schemes. 

The remainder of the paper is structured as follows: Section \ref{sec:setup} recalls the turnpike phenomenon via an illustrative example and we introduce the considered problem set-up. Moreover, we recall the basics of PCE and how  one can avoid PCE truncation errors in the LQ-setting. 
Section~\ref{sec:results} presents several examples of stochastic LQ problems, including uncertainty in the initial condition and additive stochastic disturbances. We also present an example which goes beyond the usual minimization of expected value objective. 
Finally, Section \ref{sec:summary} provides a concise summary. 

\paragraph*{Notation}
Deterministic $\{$state, input$\}$ variables are written as $x(k), u(k)$ etc., while their stochastic counterparts are written by $X(k), U(k)$. The expected value and variance are denoted as $\mbb{E}$ and $\mbb{V}$. The deterministic PCE coefficients of random variables $X(k)$, $U(k)$ are written as $\pce{x}_0$, $\pce{x}_1$, $\dots$, respectively, as $\pce{u}_0$, $\pce{u}_1$, $\dots$. $\mbb{I}_{[k_1, k_2]}$ denotes the set of positive integers $\{k_1, k_1+1, \dots, k_2\}$.

\section{Problem Set-up}\label{sec:setup}
To motivate the considered problem setting and the later stochastic examples, we first consider a motivating deterministic example taken from~\citep{Gruene13a}.
\begin{example}[Motivating example]
Consider the  following deterministic OCP
\begin{align*}
\min_{u(\cdot), x(\cdot)} \sum_{k=0}^{N-1} u^2(k) \\
\text{subject to}& \\
x(k+1) &= 2x(k)+u(k), x(0) = x_0 \\
x(k)&\in[-2,2]
\end{align*}
We solve the problem for an increasing sequence of horizons $N=3,6,...,24$. The results are presented in Fig. \ref{fig:motivating}. It can be seen that the optimal solutions all approach a neighborhood of $(0,0)$ and depart from it towards the end of the horizon. This phenomenon is known as \textit{turnpike property}~\citep{Mckenzie76,Carlson91}. 
Subsequently, we are interested in exploring the turnpike phenomenon in stochastic LQ OCPs.
\begin{figure}[b]
\begin{center}
\includegraphics[width=8.0cm]{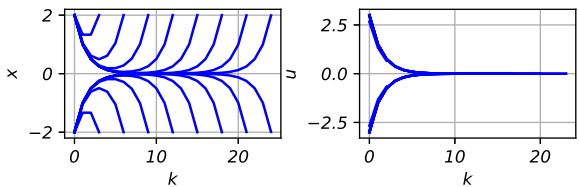}    
\caption{Solutions of the motivating example.} 
\label{fig:motivating}
\end{center}
\end{figure}

\end{example} 


\subsection{Problem Statement}
We consider stochastic LQ OCPs of the following form
\begin{subequations} \label{eq:stochOCP}
\begin{align}
\min_{U(\cdot), X(\cdot)} &~\sum_{k=0}^{N-1} \ell(X(k), U(k)) 
 \label{eq:stochOCP_obj}\\
\text{subject to} & \nonumber \\
X(k+1) & = AX(k) + BU(k) + EW(k),   
\label{eq:stochOCP_dyn}\\
X(0) &= X_0,\\
\mbb{P}[X(k) &\in \mbb{X}] \leq \varepsilon_x,  k \in \mbb{I}_{[0, N]}, \label{eq:stochOCP_Xcon}\\
\mbb{P}[U(k) &\in \mbb{U}] \leq \varepsilon_u, k \in \mbb{I}_{[0, N-1]},\label{eq:stochOCP_Ucon}
\end{align}
\end{subequations}
whereby at each time step $k$, $\randvar{x}(k) \in L^2(\Omega,\mathcal{F}_k,\mu; 
	\mathbb{R}^{n_x})$ is an $i.i.d.$ random variable on the underlying filtered probability space $(\Omega,\mathcal{F},(\mathcal{F}_k)_{k=0,\ldots,N},\mu)$, where $\Omega$ is the set of realizations, $\mathcal{F}$ is a $\sigma$-algebra, $(\mathcal{F}_k)_{k=0,\ldots,N}$ is a filtration, and $\mu$ is the probability measure.

Note that the $\sigma$-algebras $\mathcal{F}_k$ are related by the time evolution of the information, i.e. 
\[\mathcal{F}_0 \subseteq \mathcal{F}_1 \subseteq \ldots \subseteq \mathcal{F}_N \subseteq \mathcal{F}.\] 
We choose $(\mathcal{F}_k)_{k=0,\ldots,N}$ as the smallest filtration such that $\randvar{x}$ is an adapted process (which results form the evolution of the dynamics \eqref{eq:stochOCP_dyn}), i.e.\ 
\[
\mcl{F}_k = \sigma(\randvar{x}(0), \dots, \randvar{x}(k)), \quad k= 0, 1, \dots.
\]
Then, the control at time $k$ is modeled as a stochastic process which is adapted to the filtration $(\mathcal{F}_k)_k$, i.e. $\randvar{u}(k) \in L^2(\Omega,\mathcal{F}_k,\mu; 	\mathbb{R}^{n_u})$. 

The concept of stochastic filtrations can be understood as a stochastic causality requirement, i.e. the the stochastic input process $\randvar{u}(k)$ at time step $k$ may only depend on the realization of the random variables $\randvar{x}(0), \dots,\randvar{x}(k)$   up to time step $k$. Note that the influence of the noise, which is also an adapted stochastic process, is handled implicitly via the state recursion. That is, $\randvar{u}(k)$ depends on $\randvar{x}(0), \dots,\randvar{x}(k)$ and thus also on $\randvar{w}(0), \dots,\randvar{w}(k-1)$. For details on stochastic filtrations we refer to~\citep{Fristedt13a}.

	The stage cost is given by 
	\[
	\ell: L^2(\Omega,\mathcal{F},\mu; 
	\mathbb{R}^{n_x}) \times  L^2(\Omega,\mathcal{F},\mu; 
	\mathbb{R}^{n_u}) 
	\to \mbb{R}.\] 
	Indeed typical choice for $\ell$ are a combination of expected value and variance of some underlying deterministic stage cost function.  The considered dynamics are a time-discrete stochastic system subject to noise $W(\cdot)$ modelled as a stochastic process. The constraints are written as chance constraints for states and inputs, whereby the underlying sets $\mbb{X}$ and $\mbb{U}$ are assumed to be closed. Moreover, $\varepsilon_x$ and $\varepsilon_u$ specify the probabilities with which the chance constraints shall be satisfied. 

\subsection{Basics of Polynomial Chaos Expansion}
In order to obtain a tractable reformulation of the stochastic LQ OCP \eqref{eq:stochOCP} we consider the framework of Polynomial Chaos Expansions (PCE). For an in-depth introduction we refer to \citep{Sullivan2015}. The underlying idea of PCE is that random variables from some $L^2(\Omega,\mathcal{F},\mu; \mathbb{R}^{m})$ with $m \in \{n_x, n_u, n_w\}$ can be described using an appropriate basis. To this end, we consider an orthogonal  polynomial basis $\{\phi_i\}_{i=0}^\infty$ which spans $L^2(\Omega,\mathcal{F},\mu; \mathbb{R}^{m})$. 

\begin{definition}[Polynomial chaos expansion]
The polynomial chaos expansion of a real-valued random variable $\randvar{X} \in L^2(\Omega,\mathcal{F},\mu; \mathbb{R}^{n_x})$ is
\begin{equation}
\label{eq:FourierFormula}
\randvar{X} = \sum_{i = 0}^{\infty} \pce{x}_i \phi_i, \quad \text{with} \quad \pce{x}_i = \frac{\langle \randvar{x}, \phi_i \rangle}{\langle \phi_i, \phi_i \rangle},
\end{equation}
where $\pce{x}_i \in \mathbb{R}$ is called the $i$th \pce coefficient and 
\[
\langle \phi_i, \phi_j \rangle \doteq  \int_\Omega \phi_i(\omega)\phi_j(\omega)\mathrm{d}\mu(\omega),
\] 
see \citep{Sullivan2015, Xiu02}.
\end{definition}
In order to obtain a computationally tractable formulation, one has to truncate the PCE series after $L+1$ terms
\begin{equation}\label{eq:PCE_series}
{\randvar{x}} \approx \sum_{i=0}^L \pce{x}_i\phi_i.
\end{equation}
The choice of the basis polynomials can be inferred via the Wiener-Askey scheme \citep{Sullivan2015}. For example, in case of a standard Gaussian random variable one would consider Hermite polynomials as they allow modeling the Gaussian with the first two terms of the PCE series.
Moreover, in case of explicit polynomial maps---e.g., consider state transitions $\randvar{x}(k) = A^k\randvar{x}(0)$---of finite degree, one can  quantify the truncation errors arising from considering only the first $L+1$ PCE coefficients. One may also infer $L \in \mbb{N}$ such that no truncation error arises. We refer to \cite{kit:muehlpfordt18a} for details.

Finally, we remark that whenever the truncated series representation \eqref{eq:PCE_series} is exact,  the first two moments of a random variable $\randvar{X}$ can be computed in terms of PCE coefficients as follows
\begin{align*}
\mbb{E}[\randvar{x}] &= \pce{x}_0 \text{ and } \mbb{V}[\randvar{x}] = {\sum_{i=1}^L} \pce{x}_i^2 \langle \phi_i, \phi_i \rangle.
\end{align*}

\section{Simulation Study}\label{sec:results}
We consider numerical examples that show the turnpike phenomena in the stochastic setting: The first example is an extension of the motivating example but with uncertain initial condition and additive Gaussian noise.  The second example considers a linearized chemical reactor subject to non-Gaussian noise. The third example extends the second one via variance penalization in the objective.

\subsection{Scalar Dynamics with Noise}
Consider the stochastic variant of the motivating example
\begin{subequations} \label{equ:example1_dynamic}
	\begin{align}
	\randvar{x}(k+1) &= 2 \randvar{x}(k) + \randvar{u}(k) + \randvar{w}(k), \quad 
	\randvar{x}(0) = \randvar{x}_0,
	\end{align}
\end{subequations}
where $\randvar{x}_0$ denotes the initial random variable with known probability distribution $p_{\randvar{x}_0}$, $\randvar{w}(k)$ denotes system noise modeled as a white Gaussian noise such that all $W(k)$ have an identical known probability distribution $p_{\randvar{w}}$. 
We arrive at the following stochastic LQ OCP
\begin{subequations} \label{equ:example1}
	\begin{align}
	\min_{\randvar{u}(\cdot), \randvar{X}(\cdot)} \quad \sum_{k=0}^{N-1} \mbb{E}[\randvar{u}^2(k)]
	\end{align}
	\begin{align}
	\textrm{s.t.} \quad & (\ref{equ:example1_dynamic}),\; \forall k \in \mbb{I}_{[0, N-1]} \\
	& {\mbb{P}\left[ \lowbd{x} \leq \randvar{x}(k) \leq \upbd{x} \right] \geq \varepsilon_x,\; \forall k \in \mbb{I}_{[0, N]}}    \label{equ:example1_chance_constraint} \\
	& \randvar{x}_0 \sim p_{\randvar{x}_0}, \quad \randvar{w}(k) \sim p_{\randvar{w}},\; \forall k \in \mbb{I}_{[0, N-1]}
	\end{align}
\end{subequations}
where $\lowbd{x}=-2$ and $\upbd{x}=2$. 
We approximate the chance constraint as
\begin{equation}
\lowbd{x} \leq \mbb{E}[\randvar{x}(k)] \pm \lambda(\varepsilon_x) \sqrt{\mbb{V}[\randvar{x}(k)]} \leq \upbd{x}
\end{equation}
with $\lambda=\sqrt{(1+\varepsilon_x)/(1-\varepsilon_x)}$, for a derivation see~\citep{Farina13a}.

We consider  $\randvar{x}_0$ to follow a uniform distribution with the support $[\lowbd[0]{x}, \upbd[0]{x}]=[0.6, 1.4]$. The noise at time $k \in \mbb{I}_{[0,N-1]}$, $\randvar{w}(k)$, is a Gaussian distribution with mean $\mbb{E}[\randvar{w}(k)] = 0$ and variance $\mbb{V}[\randvar{w}(k)] = 0.5^2$. {Additionally,  $\varepsilon_x$ is set to 0.8 and thus we have $\lambda=3$}. 

Without the noise, considering a first-order PCE ($L =1$) for $\randvar{x}(k)$ and $\randvar{u}(k)$ with identical basis functions, exactness of the PCE representation of $\randvar{X}(k+1)$ is guaranteed since the system dynamic is linear, see \cite{kit:muehlpfordt18a}. Therefore, including noise, the  PCE dimension needed for an exact representation is determined by the horizon $N$. More precisely, $L=N+1$,  where two PCE terms are induced by the uncertainty of the initial condition and the rest is caused by the noise. The PCE basis and coefficients read
\begin{subequations}
	\begin{gather}
	\randvar{x}(k)   = \sum_{i=0}^{1} \pce{x}_i(k) \phi_i + \sum_{i=0}^{N-1} \pce{x}_i^{w}(k) \phi_i^w\\
	\randvar{u}(k)   = \sum_{i=0}^{1} \pce{u}_i(k) \phi_i + \sum_{i=0}^{N-1} \pce{u}_i^{w}(k) \phi_i^w\\
	\randvar{w}(k)   = \sum_{i=0}^{N-1} \pce{w}_i^{w}(k) \phi_i^w\\
	\randvar{x}_0 = \sum_{i=0}^{1} \pce{x}_{0i} \phi_i
	\end{gather}	
	with
	\begin{gather}
	\phi_0 = 1, \; \phi_1 = \xi - 0.5, \; \phi_i^w = \theta_i, i \in \mbb{I}_{[0,N-1]} \label{equ:pce_basis_example1} \\
	\pce{w}_k^{w}(k) = \sqrt{\mbb{V}[\randvar{w}(k)]}, \; \pce{w}_i^{w}(k) = 0, i\in \mbb{I}_{[0,N-1]} \setminus \{k\}
	\end{gather}
\end{subequations}
where $\xi$ is a standard uniformly distributed random variable while $\theta_i$ are independent Gaussian random variables. Additionally, $\pce{x}_{00}=(\lowbd[0]{x}+\upbd[0]{x})/2$ and $\pce{x}_{01}=\upbd[0]{x}-\lowbd[0]{x}$, since $\randvar{x}_0$ is uniformly distributed.

\subsection*{Expressing Stochastic Filtrations in the PCE Framework}

{Stochastic modeling via adapted filtrations expresses the idea that the noise $\randvar{w}(k)$ at time $k$  influences the next state $\randvar{x}(k+1)$ and input $\randvar{u}(k+1)$ and the subsequent time instances $i > k$ but not at $i \leq k$.} In terms of PCE representation this implies that
 \begin{multline}
\begin{bmatrix} \randvar{z}(0) \\ \randvar{z}(1) \\ \randvar{z}(2) \\ \vdots \\ \randvar{z}(M) \end{bmatrix} = 
\begin{bmatrix} \pce{z}_0(0) & \pce{z}_1(0)\\   \pce{z}_0(1) & \pce{z}_1(1) \\   \pce{z}_0(2) & \pce{z}_1(2) \\
 \vdots & \vdots \\
 \pce{z}_0(M) & \pce{z}_1(M) 
 \end{bmatrix}  \begin{bmatrix} \phi_0 \\ \phi_1 \end{bmatrix}
 +
\pce{Z}^w
\begin{bmatrix} \phi_0^w \\ \phi_1^w\\\phi_2^w \\ \vdots \\ \phi_{M-1}^w \end{bmatrix}
\label{equ:example1_causal}
\end{multline}
with
\[
\pce{Z}^w \doteq \begin{bmatrix}
0 & 0 & \cdots & 0\\
	\pce{z}_0^w(1)   & 0          & \cdots   & 0   \\
	\pce{z}_0^w(2)   & \pce{z}_1^w(2)    & \cdots   & 0  \\
	\vdots    & \vdots     & \ddots   & \vdots  \\
	\pce{z}_0^w(M) & \pce{z}_1^w(M)  & \cdots\  & \pce{z}_{M-1}^w(M)
	\end{bmatrix} \in \mbb{R}^{(M+1)\times M},
\]
and where, respectively, $\randvar{z} \in \{\randvar{x}, \randvar{u}\}$ and $\pce{z} \in \{\pce{x}, \pce{u}\}$ are placeholders and $M = N$ for $\randvar{X}$ and $M = N-1$ for $\randvar{U}$.  

\subsection*{Results}
\begin{figure}[t]
	\begin{center}
		\includegraphics[width=8.0cm]{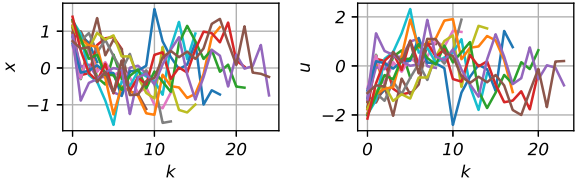}    
		\caption{16 sample realization trajectories of optimal solutions to  OCP \eqref{equ:example1_causal}.} 
		\label{fig:example1_sample}
	\end{center}
\end{figure}

 To illustrate turnpike behavior, we solve the stochastic OCP (\ref{equ:example1}) as given above over different optimization horizons $N=3,6,...,24$ using \texttt{PolyChaos.jl} \citep{tudo:muehlpfordt19c}. Fig. \ref{fig:example1_sample} shows the trajectories of the optimal solutions for a total of 16 realizations of the uncertainties. At first glance, the realizations of the solutions are noisy and appear not to exhibit the turnpike property.

Hence, in order to uncover the turnpike, we consider the trajectories of the PCE coefficients. Note that each PCE basis $\phi_i^w$ induced by the noise is a standard Gaussian distributed random variable, the sum $\sum_{i=0}^{k-1} \pce{z}_i^{w}(k) \phi_i^w, \pce{z} \in \{\pce{x}, \pce{u}\}$ is equal to a new Gaussian distributed random variable with zero mean and variance. Hence, for the sake of simplified illustration, we consider
\begin{subequations}
\begin{equation}
\pce{z}_\Sigma^w(k) \phi_\Sigma^w = \sum\nolimits_{i=0}^{k-1} \pce{z}_i^{w}(k) \phi_i^w
\end{equation}
with
\begin{equation}
\phi_\Sigma^w = \theta_\Sigma,\quad \pce{z}_\Sigma^w(k) = \sqrt{\sum\nolimits_{i=0}^{k-1} (\pce{z}_i^w(k))^2}.
\end{equation}
\end{subequations}

\begin{figure}[t]
	\begin{center}
		\includegraphics[width=8.0cm]{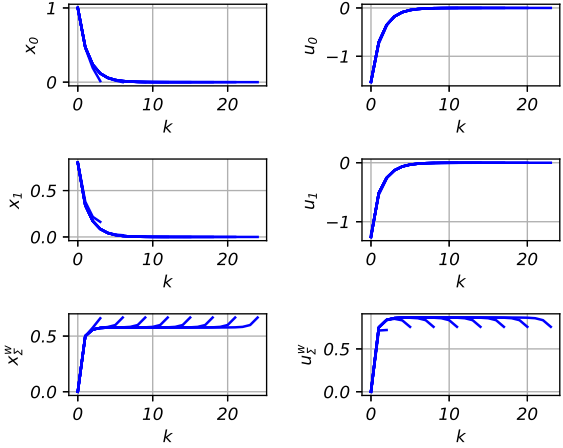}    
		\caption{PCE coefficients solutions to  OCP \eqref{equ:example1_causal}.} 
		\label{fig:example1_turnpike_pce}
	\end{center}
\end{figure}

Here $\phi_\Sigma$ is a random variable with standard Gaussian distribution (zero mean). Therefore,  instead of $\pce{z}_i^w(k), i\in \mbb{I}_{[0,N-1]}$, only one PCE coefficient $\pce{z}_\Sigma^w(k)$ suffices to represent the uncertainty caused by noise. Note that this transformation is used only for  illustration priposes and not in the underlying computation. 

\begin{figure}
	\begin{center}
		\includegraphics[width=5.4cm]{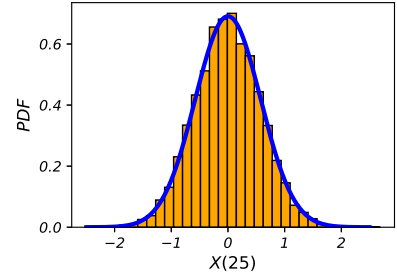}    
		\caption{Comparison between PDF of the optimal steady state and the histogram of state trajectories at $k = 25 $.} 
		\label{fig:example2_distribution_stat}
	\end{center}
\end{figure}

Using the PCE reformulation detailed above, Fig. \ref{fig:example1_turnpike_pce} illustrates that the turnpike phenomenon occurs in terms of PCE coefficients. Actually, the turnpike property of PCE coefficients suggests that the optimal steady-state is a random variable with  stationary distribution. This distribution can be calculated from the optimal steady-state problem formulated via PCE. Doing so, i.e. solving 
\begin{align*}
\min_{\bar{\randvar{X}}, \bar{\randvar{U}}} &~\bar{\randvar{U}}^2 \\
\text{subject to}&\\
\bar{\randvar{x}} &= 2 \bar{\randvar{x}} + \bar{\randvar{u}} + \bar{\randvar{w}} \\
\lowbd{x} &\leq \mbb{E}[\bar{\randvar{x}}] \pm \lambda(\varepsilon_x) \sqrt{\mbb{V}[\bar{\randvar{x}}]} \leq \upbd{x},
\end{align*}
we obtain the stationary distribution depicted in Fig. \ref{fig:example2_distribution_stat}. This figure also shows the histogram at $k =25$ obtained from sampling $10^4$ realizations of the uncertainty from the optimal PCE solution for $N=50$. As one can see, the behavior in the middle of the horizon corresponds to the solution obtained for the steady state problem. 

\begin{figure}
	\begin{center}
		\includegraphics[width=7.5cm,trim=60 40 80 80,clip=true]{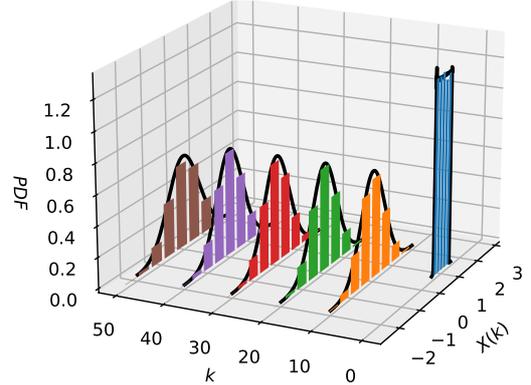}    
		\caption{Comparison of PDF and histogram of $10^4$ samples for  OCP \eqref{equ:example1_causal} with $N=50$.} 
		\label{fig:example1_distribution}
	\end{center}
\end{figure}

The time evolution of the state histograms and the distributions obainted via PCE is shown in Fig. \ref{fig:example1_distribution} for $k =0,10,20,30,40,50$. It is not surprising that the histograms follow the calculated PDF perfectly and the state $\randvar{x}(k)$ keeps the same distribution in the middle of trajectory. This illustrates the turnpike phenomenon in the distributions. 


\subsection{Stochastic LQ OCP for a CSTR}
\begin{figure}
	\begin{center}
		\includegraphics[width=7.9cm]{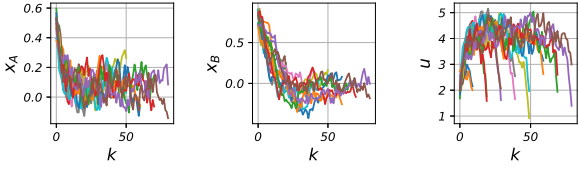}   
		\caption{Sample trajectories for the OCP with \eqref{eq:CSTR1}.} 
		\label{fig:example2_sample}
	\end{center}
\end{figure}

\begin{figure}
	\begin{center}
		\includegraphics[width=7.9cm]{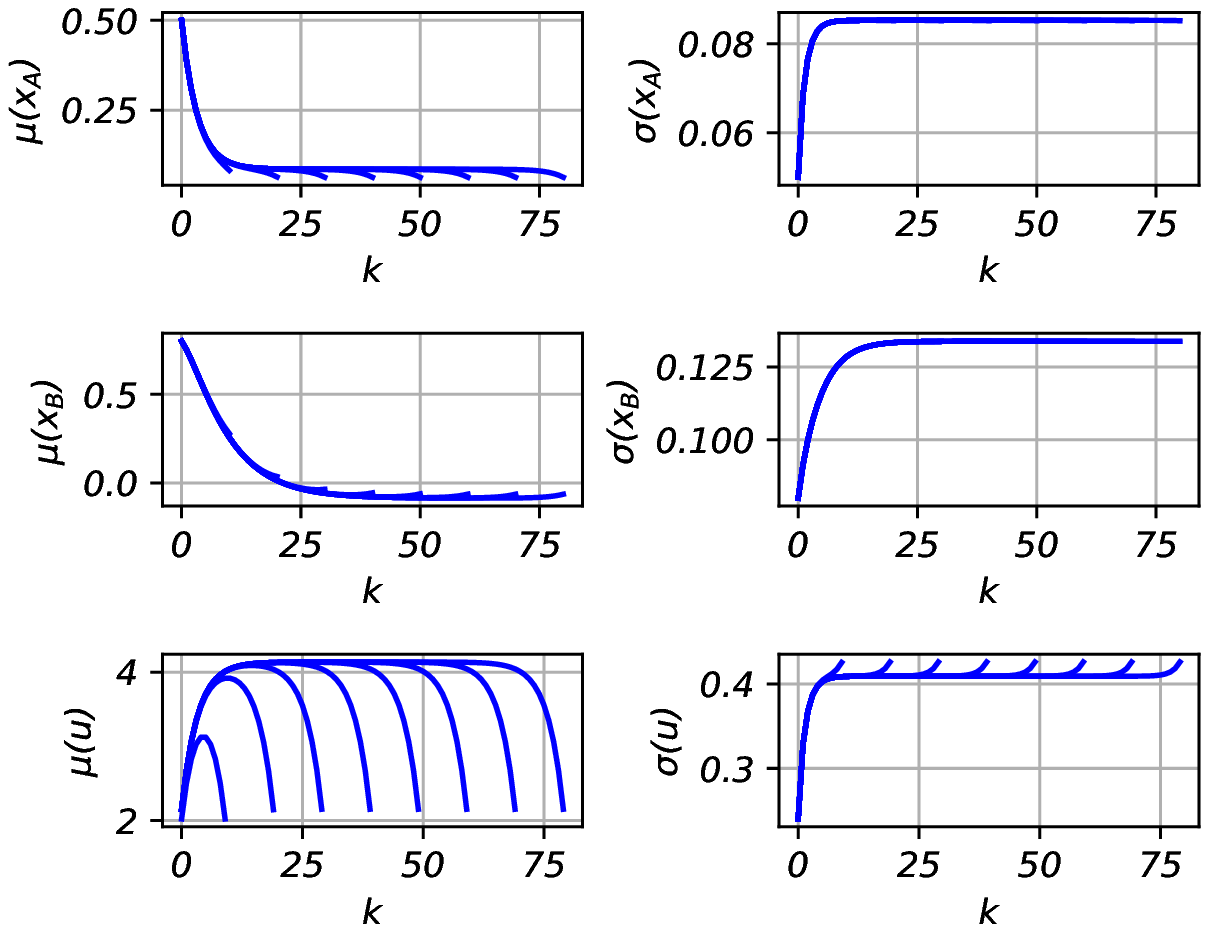}    
		\caption{Mean and variance obtained from the PCE coefficients for the OCP with \eqref{eq:CSTR1}.} 
		\label{fig:example2_turnpike_pce}
	\end{center}
\end{figure}

\begin{figure}
	\begin{center}
		\includegraphics[width=7.4cm,trim=60 40 80 80,clip=true]{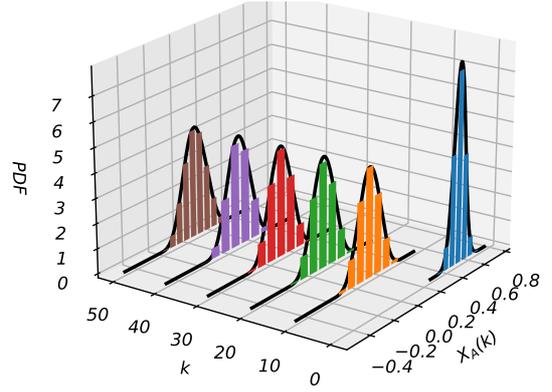}    
		\caption{Comparison of PDF and histogram of $10^4$ samples for the OCP with \eqref{eq:CSTR1} $N=50$.} 
		\label{fig:example2_distribution}
	\end{center}
\end{figure}

As a second example, we consider a linearized CSTR reactor which appeared in several papers such as \citep{kit:zanon18a}.
The expected value quadratic stage cost and the linear discrete-time system with noise read
\begin{subequations}\label{eq:CSTR1}
\begin{align}
\ell(\randvar{x},\randvar{u}) &=  \mbb{E}[\randvar{x}_A + 0.1 \randvar{u}^2 + 24\randvar{x}_B - 0.5\randvar{u}] \\
f(\randvar{x},\randvar{u}) &= \begin{bmatrix} 0.76 & 0\\ 0.12 & 0.88 \end{bmatrix} \begin{bmatrix} \randvar{x}_A\\ \randvar{x}_B \end{bmatrix} + \begin{bmatrix} \phantom{-}0.005\\ -0.005 \end{bmatrix} \randvar{u} + \begin{bmatrix} \randvar{w}_A\\ \randvar{w}_B \end{bmatrix}
\end{align}
\end{subequations}
where $\randvar{w}_A,\randvar{w}_B$ are modeled as independent uniformly distributed random variables with support $[-0.1,0.1]$. The initial state $\randvar{x}(0)$ is  a Gaussian distributed random vector with known mean $(0.5,0.8)$ and variance $(0.05^2,0.08^2)$. We solve the stochastic OCP via PCE over different horizons $N=10,20,...,80$. We obtain the trajectories of the optimal solutions for a total $16$ realization sequences of the uncertainties, see Fig. \ref{fig:example2_sample}. As the dimension of PCE is quite large, we plot the first two moments instead of PCE coefficients, i.e. mean and the variance of state and input random variables, see Fig. \ref{fig:example2_turnpike_pce}.  Similar to the previous example, the trajectories of mean and variance exhibit the turnpike property. Similar to before,  Fig. \ref{fig:example2_distribution} depicts the histograms of the state $\randvar{x}_A$ at $k = 0,10,20,30,40,50$ for the considered realizations and the PDF obtained via PCE, see Fig. \ref{fig:example2_distribution}.

\begin{figure}
	\begin{center}
		\includegraphics[width=7.7cm]{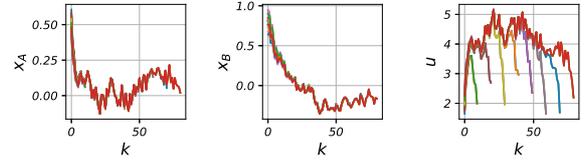}    
		\caption{System response and corresponding input for the OCP with \eqref{eq:CSTR1}:  different initial conditions and horizons with one identical noise realization.} 
		\label{fig:example2_fixed_sample}
	\end{center}
\end{figure}

As an additional means of assessing the turnpike phenomenon via simulation, we consider the following numerical setting: 
\begin{itemize}
\item Compute a random realization of the disturbance $\randvar{W}(k)$,  for $k \in \mbb{I}_{[0,M]}$ denoted as $\{\randvar{W}(k;\omega_k)\}$
\item Pick horizons $N_1, ..., N_s \leq M$ and corresponding realizations of the initial condition $X_0(\omega_1), \dots, X_0(\omega_s)$.
\item For all horizons $N_1, \dots, N_s$ and the initial condition $X_0(\omega_1), \dots, X_0(\omega_s)$, simulate the response of the dynamics under the optimal input policy, while the disturbance sequence is fixed to $\{\randvar{W}(k;\omega_k)\}$ (or a subpart thereof). 
\end{itemize}
The results of this numerical experiment are depicted in Figure \ref{fig:example2_fixed_sample}.
As one can see, all the trajectories approach the same solution after some time. One can understand this solution as a time-varying path of a stationary turnpike solution, whose shape is governed by the considered disturbance sequence. Observe the difference to Figure \ref{fig:example2_sample}, wherein for each trajectory a different disturbance realization sequence is considered.

\subsection{Stochastic LQ OCP with Variance Penalization}

\begin{figure}[t]
	\begin{center}
		\includegraphics[width=8.0cm]{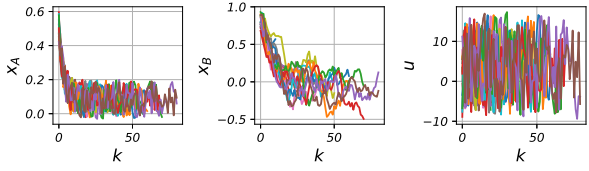}    
		\caption{Sample trajectories  for the OCP with \eqref{eq:CSTRII}.} 
		\label{fig:example3_punish_sample}
	\end{center}
\end{figure}

\begin{figure}[t]
	\begin{center}
		\includegraphics[width=8.0cm]{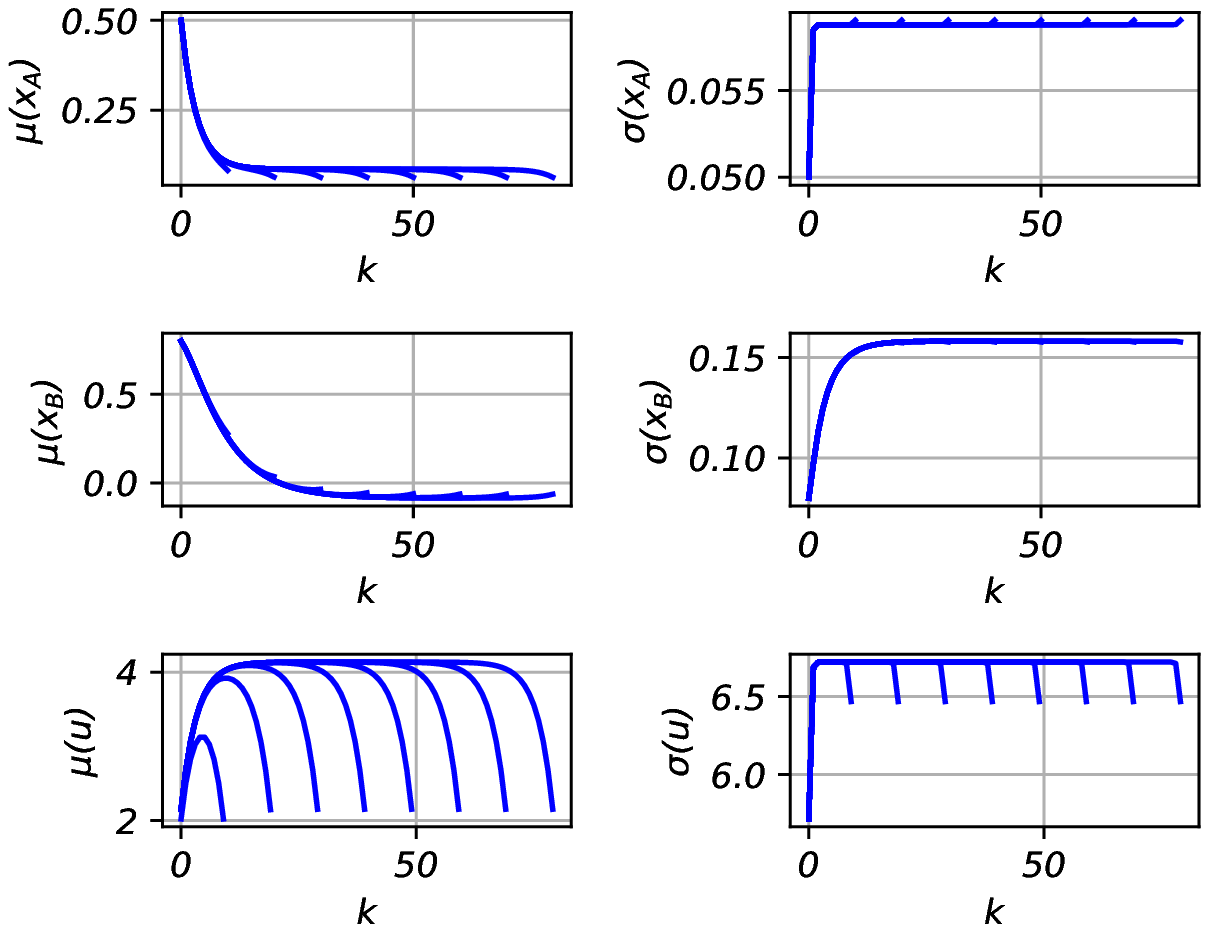}    
		\caption{Mean and variance obtained from the PCE coefficients for the OCP with \eqref{eq:CSTRII}.} 
		\label{fig:example3_punish_turnpike_pce}
	\end{center}
\end{figure}

\begin{figure}[t]
	\begin{center}
		\includegraphics[width=8.0cm,trim=60 40 80 80,clip=true]{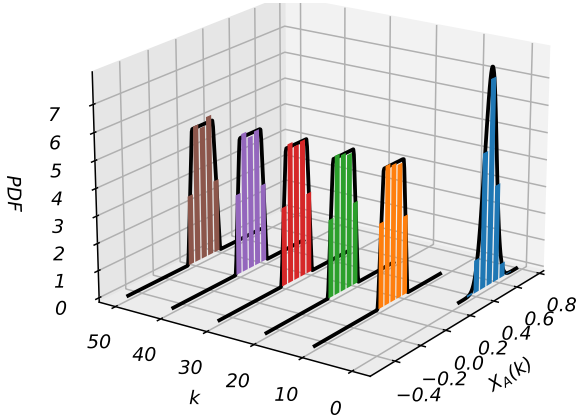}    
		\caption{Comparison of PDF and histogram of $10^4$ samples for the OCP with \eqref{eq:CSTRII} $N=50$.} 
		\label{fig:example3_punish_distribution}
	\end{center}
\end{figure}

What could we do if we want to get a optimal steady-state with a narrow distribution, or in other words, with small variance? Involving variance penalization in the objective function is one option. We consider the previous  example augmented with a penalty of the variance of the state $\randvar{x}_A$ in the stage cost
\begin{equation}\label{eq:CSTRII}
\ell (\randvar{x},\randvar{u}) = \mbb{E}[\randvar{x}_A + 0.1 \randvar{u}^2 + 24\randvar{x}_B - 0.5\randvar{u}] + \gamma\mbb{V}[\randvar{x}_A]
\end{equation}
Here we choose $\gamma=10^4$ and solve the stochastic OCP via PCE over different horizons $N=10,20,...,80$. The trajectories of the optimal solutions for a total $16$ realization sequences of the uncertainties are shown in Fig. \ref{fig:example3_punish_sample}. Observe that the state $x_A$ is much less effected by the noise. It can also be seen in Fig. \ref{fig:example3_punish_turnpike_pce} that the variance of state $\randvar{x}_A$ is smaller than in Example 2, while the variance of state $\randvar{x}_B$ and the variance of the input $\randvar{u}$ increase. Fig. \ref{fig:example3_punish_distribution} also shows that the optimal steady state $X_A$ has a narrower distribution.\\

\section{Summary} \label{sec:summary}
This paper has conducted a simulation study on turnpike properties in stochastic OCPs. It has presented three examples of stochastic LQ OCPs all of which exhibit  the turnpike phenomenon. Indeed, the examples demonstrate that turnpike phenomena can be observed in different contexts:
\begin{itemize}
\item in terms for statistical moments (or PCE coefficients which can be mapped to moments),
\item in terms of probability distributions of state and input vairables staying close to their optimal stationary distributions, and
\item in terms of the realization trajectories staying close to an orbit defined by the noise realization. 
\end{itemize}
Moreover, our simulation study demonstrates that beyond the usual minimization of expected values, the turnpike phenomenon is also present in combination of expected value and min-variance objectives.  

While this note did merely present simulation results, there is a clear prospect of extending the established notations of turnpike properties to stochastic OCPs and corresponding stochastic MPC formulations. Yet, at this stage, there is also an evident need for analytic results to understand the turnpike phenomenon in stochastic settings.

\def\bibfont{\tiny}

\bibliography{literature_latin1}            

\end{document}